\documentclass[twocolumn]{autart}   
\usepackage{graphicx} 
\usepackage[a4paper,tmargin=.75in, lmargin=.65in, bmargin=.95in, rmargin=.65in]{geometry}
\usepackage{comment} 
\usepackage{amssymb,amsmath, mathrsfs,mathtools}
\usepackage{setspace}
\usepackage{tabularray}
\usepackage{multirow}
\usepackage{array}
\usepackage{bigints}
\usepackage{xurl}
\usepackage{wrapfig}
\usepackage[authoryear,round]{natbib}
\usepackage[dvipsnames]{xcolor}
\usepackage[colorlinks,allcolors=RoyalBlue]{hyperref}

\begin{document}

\begin{frontmatter}

\title{Simulation-based Approach for Fast Optimal Control of a Stefan Problem with Application to Cell Therapy} 

\vspace{-.7cm}

\author[MIT]{Prakitr Srisuma}\ead{prakitrs@mit.edu},
\author[MIT]{George Barbastathis}\ead{gbarb@mit.edu}, 
\author[MIT]{Richard D. Braatz}\ead{braatz@mit.edu}

\thanks[footnoteinfo]{Corresponding author: Richard D. Braatz}

\thanks[footnoteinfo2]{This manuscript is an extension of a paper published in the Proceedings of the American Control Conference \citep{Srisuma_2024_OCPstefan}.}

\address[MIT]{Massachusetts Institute of Technology, Cambridge, MA 02139, USA}


\vspace{-.3cm}
\begin{abstract}
This article describes a new, efficient way of finding control and state trajectories in optimal control problems by reformulation as a system of differential-algebraic equations (DAEs). The optimal control and state vectors can be obtained via simulation of the resulting DAE system with the selected DAE solver, eliminating the need for an optimization solver. Our simulation-based approach is demonstrated and benchmarked against various optimization-based algorithms via four case studies associated with the optimization and control of a Stefan problem for cell therapy. The simulation-based approach is faster than every optimization-based method by more than an order of magnitude while giving similar/better accuracy in all cases. The solution obtained from the simulation-based approach is guaranteed to be optimal provided that at least one constraint or algebraic equation resulting from the reformulation remains active at all times. The proposed technique offers an efficient and reliable framework for optimal control, serving as a promising alternative to the traditional techniques in applications where speed is crucial, e.g., real-time online model predictive control.
\end{abstract}

\end{frontmatter}
\endNoHyper

\section{Introduction} \label{Introduction}
A Stefan problem describes the evolution of a moving interface during phase change, e.g., freezing and melting \citep{Carslaw_1959_Conduction,Bird_2002_transpheno}. Different Stefan problem formulations have been applied to study various industrial and natural systems, including polymorphous materials \citep{Tao_1979_Polymorphous}, steel casting \citep{Hill_1994_CastingInterface,Petrus_2010_Feedback2Phase}, biological tissue \citep{Rabin_1995_CellThawingNonIdeal,Rabin_1997_CellThawingNonIdeal}, alloy formation \citep{Planella_2019_BinaryAlloy,Planella_2021_BinaryAlloy}, glaciation \citep{Mikova_2017_glaciation}, phase change materials \citep{Brezina_2018_PCM}, cryopreservation \citep{Mohit_2020_Cryopreservation}, and cell therapy \citep{Srisuma_2023_1DCellThawingEstimation}. Numerical techniques have been developed for implementing and simulating the Stefan problems  \citep{Meyer_1971_numer2phase,Veraldi_2008_Parameters,Kurbatova_2019_numer2phase,Gusev_2021_FVM2Phase,Srisuma_2023_1DCellThawingEstimation}.

Optimal control of Stefan problems has been extensively investigated and proven useful for various industrial applications over the past few decades, with many different objective functions, constraints, and controls (manipulated variables) considered. Some examples are the control of a heating process to satisfy constraints on the heating speed and thermoelastic stress by manipulating the furnace temperature \citep{Roubíček_1986_OC_CircularPlate}, the stabilization of the moving boundary and temperature/concentration fields by varying the heat flux \citep{Pawlow_1987_OC_StablizeField}, the maximization of the amount of melted solid in a melting process via controlling the heat flux \citep{Neto_1994_OC_MaxMelt}, the control of the moving boundary to follow the desired path in a solidification process by manipulating the wall temperature \citep{Hinze_2007_OC_Interface}, the control of the water level in a drainage basin by varying the discharge velocity  \citep{Miyaoka_2008_OC_Basin}, the optimization of a thin-film drying process via manipulating the air temperature \citep{Mesbah_2014_Drying}, and the minimization of the metallurgical length deviation in steel casting by controlling the boundary heat flux \citep{Chen_2019_OC_MLlength}.

All the aforementioned studies rely on optimization algorithms/solvers to obtain the optimal control trajectories. The widely used approach is to replace the time-varying control vector by a finite-dimensional parameterization (e.g., a spline) and carry out numerical discretization to transform the dynamic optimization into a nonlinear algebraic program \citep[for example]{Kishida_2013_OptimalControl,Scott_2018_dFBA,Nolasco_2020_OptimalControl}. This overall approach has many variations, including methods that sequentially switch between solving the numerical discretization of the underlying partial differential-algebraic equations (PDAEs) and running an algebraic optimizer, or simultaneously solving a single large sparse optimization with the numerical discretization equations as explicit constraints \citep{Neuman_1974_simultaneous,Tsang_1975_simultaneous,Mellefont_1978_sequential,Sargent_1978_sequential,Biegler_2007_NLP,Nolasco_2020_OptimalControl}.
Numerous optimal control algorithms have been developed for efficient solutions to facilitate real-time applications such as model predictive control (e.g., see detailed discussions by \citet{Rodrigues_2014_OCoverview} and \citet{Nolasco_2020_OptimalControl}). Alternatively, \citet{Berliner_2022_battery} showed that it is possible to reformulate some optimal control problems as a mixed continuous-discrete system of index-1 PDAEs. The PDAE system is then solved numerically by feeding the differential-algebraic equations (DAEs) obtained by spatial discretization into a DAE solver. In this approach, the optimal control vector is obtained directly from the DAE solver, without using any optimization solver, resulting in a highly computationally efficient solution to the optimal control problem. Recently, this approach was used in the optimal control of a Stefan problem \citep{Srisuma_2024_OCPstefan}. 


Although optimal control of Stefan problems has been explored for a wide range of processes, applications to cell therapy are lacking. Previous studies have shown that accurate prediction, control, and optimization of cryopreservation and cell thawing can improve the viability and quality of the resulting cells, which directly benefits cell therapy \citep{Shinsuke_2008_ControlledHeating,Jang_2017_CryopOverview,Baboo_2019_ThawingRates,Hunt_2019_TechnicalCondThawing,Bojic_2021_Winter,Cottle2022,Uhrig_2022_ThawingOpt}. These benefits and case studies therefore motivate the development of an efficient Stefan problem-based optimal control algorithm for cell therapy.

This article describes an optimal control algorithm via DAE reformulation and simulation, referred to as the simulation-based approach, for cell therapy applications. The main contributions of this work are to
\vspace{-5pt}
\begin{enumerate}
\item derive the simulation-based approach for optimal control of a Stefan problem applied to the cell thawing process in cell therapy; 
\item generalize the simulation-based approach to both index-1 and high-index DAE systems;
\item demonstrate and benchmark the simulation-based algorithm against various optimization-based optimal control algorithms;
\item apply the proposed approach to efficiently and accurately solve several practical problems associated with the optimization and control of cell thawing; and
\item discuss its limitation and condition required for the approach to provide an optimal solution.
\end{enumerate}

This article is organized as follows. Section \ref{sec:ModelFormulation} describes the cell thawing system and summarizes the mechanistic model and equations. Section \ref{sec:OptimalControl} discusses various techniques for solving optimal control problems and introduces the simulation-based approach. Section \ref{sec:CaseStudies} demonstrates the simulation-based approach via case studies. Finally, Section \ref{sec:Conclusion} summarizes the study.

\section{Stefan Problem and Cell Thawing} \label{sec:ModelFormulation}
\begin{figure}
\centering
    \includegraphics[scale=0.46]{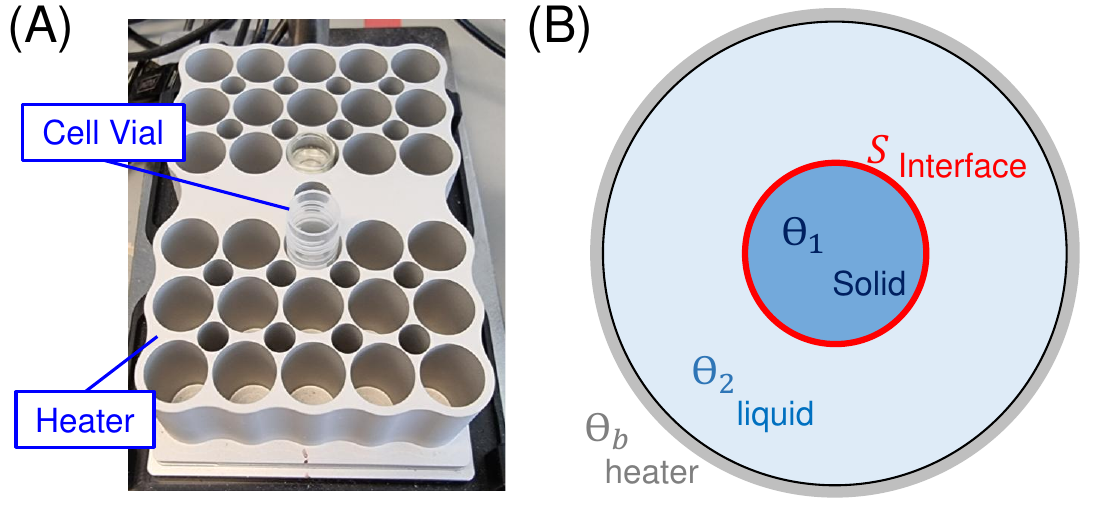}

    \caption {(A) In cell thawing, the vial containing biological cells frozen in ice is heated by the heater. (B) A schematic diagram showing the one-dimensional Stefan problem in a cylindrical coordinate system. The moving interface position is $S$. The solid temperature is $\Theta_1$. The liquid temperature is $\Theta_2$. The heater temperature is $\Theta_b$. All variables are written in dimensionless form.} 
    \label{fig:Schematic}      
\end{figure}

The system used for the demonstration of our optimal control technique in this article concerns cell thawing, which is a process used in cell therapy before cells are introduced into the patients (Fig.~\ref{fig:Schematic}). During thawing, energy is continuously supplied by a heater to thaw the material in a vial. A mechanistic model of cell thawing can be formulated as a Stefan problem
\citep{Srisuma_2023_1DCellThawingEstimation}. Heat transfer in the solid and liquid domains can be described by the energy balance equation,
\begin{equation}
\label{eq:SimplifiedEnergyEquation}
\frac{1}{\alpha}\frac{\partial T}{\partial t} = \frac{\partial^2 T}{\partial r^2}+\frac{1}{r} \frac{\partial T}{\partial r},
\end{equation}
where $T$ is the temperature, $r$ is the radial direction, $t$ is time, and $\alpha$ is the thermal diffusivity. Heat transfer associated with thawing at the moving solid-liquid interface is governed by the Stefan conditions
\begin{gather} 
\rho \Delta H_f \dfrac{ds}{dt} = k_1\dfrac{\partial T_1}{\partial r}-k_2\dfrac{\partial T_2}{\partial r} \label{eq:Stefan1},\\
T_1=T_2=T_m, \label{eq:Stefan2}
\end{gather}
where $s$ is the interface position, $\Delta H_f$ is the latent heat of fusion, $T_m$ is the melting point, $\rho$ is the density, $k$ is the thermal conductivity, and the subscripts 1 and 2 denote the solid and liquid phases, respectively. The above equations are nondimensionalized and discretized using the finite difference scheme with the method of lines, with appropriate boundary conditions, resulting in
\begingroup
\allowdisplaybreaks
\begin{align}
&\dfrac{d\Theta_1}{d\tau} = f_1(\Theta_1, \Theta_2, S), \label{eq:model1}\\
&\dfrac{d\Theta_2}{d\tau} = f_2(\Theta_1, \Theta_2 , \Theta_b, S), \label{eq:model2}\\
&\dfrac{dS}{d\tau} = f_3(\Theta_1, \Theta_2, S), \label{eq:model5} \\
&(\Theta_1)_{n} = (\Theta_2)_{0} = 0 , \label{eq:model6} 
\end{align}
with the initial conditions
\begin{gather}
\Theta_1(\tau_0) = \Theta_2(\tau_0)  = 0, \label{eq:model7} \\
S(\tau_0) = 1, \label{eq:model8} 
\end{gather}
\endgroup
where $\Theta_1 \in \mathbb R^n$ collects the solid temperature $(\Theta_1)_i$ for $i=0,\dots{},n-1$; $\Theta_2 \in \mathbb R^n$ collects the liquid temperature $(\Theta_2)_j$ for $j=1,\dots{},n$;  $S \in \mathbb R^1$ is the interface position; $f_1 \in \mathbb R^n$, $f_2 \in \mathbb R^n $, $f_3 \in \mathbb R^1$ are the nonlinear functions; $n$ is the number of grid points in each domain (set to 20); $\tau \in [\tau_0, \tau_f]$ is the dimensionless time; $\tau_0$ is the initial time; and $\tau_f$ is the final time. Here \eqref{eq:model1} and \eqref{eq:model2} are derived from \eqref{eq:SimplifiedEnergyEquation}, while \eqref{eq:model5} and \eqref{eq:model6} correspond to \eqref{eq:Stefan1} and \eqref{eq:Stefan2}, respectively. We refer to \citet{Srisuma_2023_1DCellThawingEstimation} for the detailed derivation of all equations, parameter values, and model validation.

In dimensionless form, the model consists of three main parts that describe the evolution of the solid temperature \eqref{eq:model1}, the liquid temperature \eqref{eq:model2}, and the interface position \eqref{eq:model5}, respectively. The system is initially in the solid state, which corresponds to $S=1$; i.e., the interface position is at the outer boundary. As the thawing progresses, the interface position decreases and eventually becomes 0, indicating complete melting. The heater temperature is $\Theta_b$, which can be fixed or varied with time. The maximum dimensionless temperature is 1, which is equivalent to $37\,^\circ$C, the highest temperature recommended for cell thawing \citep{Baboo_2019_ThawingRates,Uhrig_2022_ThawingOpt}. The minimum dimensionless temperature is 0, corresponding to the melting point of $-2\,^\circ$C. 

Finally, since the temperature varies spatially, we define the average temperature $\Theta_{\textrm{avg}}$ as
\begin{equation} \label{eq:Tavg_numer}
        \Theta_{\textrm{avg}} = \sum_{j=1}^{n} \dfrac{1}{n}{(\Theta_2)_j}.
    \end{equation}
In this case, only the liquid temperature $\Theta_2$ is considered because the solid temperature $\Theta_1$ is equal to the melting point, which does not change with time. Note that, as our problem is defined in the radial direction, the average temperature over the cross section could also be used instead of \eqref{eq:Tavg_numer} for more accurate results. In any case, the implementation and algorithms presented later in this article are still valid for both definitions.

\section{Optimal Control} \label{sec:OptimalControl}
\subsection{Optimal control formulation} \label{sec:OC_Formulation}
The optimal control problem for cell thawing is
\begin{align}\label{eq:OptControlGeneral}
&\begin{aligned}
\min_{\Theta_b(\tau)} \   &\mathcal{M}(\Theta_1(\tau_f),\Theta_2(\tau_f),S(\tau_f)) +\hspace{1in} \\ &\int_{\tau_0}^{\tau_f}\mathcal{L}(\Theta_1(\tau),\Theta_2(\tau),S(\tau),\Theta_b(\tau),\tau) d\tau
\end{aligned}\\
&\textrm{s.t. Equations \eqref{eq:model1}--\eqref{eq:model8}}, \nonumber \\
&\quad \ \ \, 0 \leq \Theta_b(\tau) \leq 1 \label{eq:heatertemp_constraint}.
\end{align} 
The control (aka decision variable, manipulated variable, input) of this optimal control problem is the heater temperature $\Theta_b(\tau)$. The mechanistic model \eqref{eq:model1}--\eqref{eq:model8} describes the physics of cell thawing and thus serves as the constraints of the optimization. The lower and upper bounds on the heater temperature are represented by \eqref{eq:heatertemp_constraint}. 

\subsection{Optimization-based approach} \label{sec:Opt_based}
A detailed discussion on numerical algorithms for solving optimal control problems can be found in \citet{Nolasco_2020_OptimalControl}. The common technique for solving optimal control problems is to discretize the partial differential equation (PDE) and ordinary differential equation (ODE) constraints, parameterize the time-varying control vector, and solve the resulting optimization numerically with a proper optimizer. This approach relies on the convergence of the optimization algorithm/solver to obtain the optimal control trajectory, and hence we denote this approach as the \textit{optimization-based} approach. In this article, six different optimization-based implementations that have been used widely in optimization and optimal control applications, including a variety of software tools and programming languages, are considered:
\vspace{-5pt}
\begin{itemize}
    \item opt\_Ipopt: The control vector is parameterized, and the resulting optimization is numerically solved by IPOPT \citep{Wachter_2006_IPOPT}, an open-source optimization solver for large-scale nonlinear optimization that has been implemented in various nonlinear and optimal control software packages. The IPOPT solver is employed in MATLAB via the OPTI Toolbox \citep{OPTI_2012}. The system of ODEs is integrated using \texttt{ode15s}. 
    
    \item opt\_fmincon: The control vector is parameterized, and the resulting optimization is numerically solved by \texttt{fmincon}, a built-in optimization solver for constrained nonlinear optimization in MATLAB. The system of ODEs is integrated using \texttt{ode15s}. 
    
    \item opt\_pfmincon: The implementation is identical to opt\_fmincon except that the parallel computing option for \texttt{fmincon} is turned on.
    
    \item opt\_sCasADi: The optimal control problem is solved using CasADi \citep{Casadi_2019}, a well-established open-source tool that has been widely used for nonlinear optimization, optimal control, and model predictive control. The problem is solved symbolically by the direct single shooting method via CasADi v3.6.3, with the code directly modified from the given example pack.
    
    \item opt\_mCasADi: The implementation is similar to opt\_sCasADi but uses the direct multiple shooting method instead.

    \item opt\_Gekko: The optimal control problem is solved using GEKKO \citep{GEKKO}, a Python package for optimization and machine learning.    
\end{itemize}

\subsection{Simulation-based approach} \label{sec:Sim_based}
The approach in this article is inspired by a recent article on the optimal control of lithium-ion batteries \citep{Berliner_2022_battery}.
The approach was motivated by the observation that the optimal control trajectories for that application moved from one active inequality constraint to another over time. That information was used to transform the optimal control problem into a mixed continuous-discrete system of index-1 DAEs, in which existing software can be used to transition between active inequality constraints. This \textit{simulation-based} technique eliminates the need for an optimization solver, resulting in a much more computationally efficient solution. No control vector parameterization is needed.

\subsubsection{Solution of high-index DAEs}
In this work, we extend the simulation-based technique to handle high-index DAE systems and apply the approach to solve multiple cell thawing problems. Various numerical algorithms for solving index-1 DAEs, including the implementation used in \citet{Berliner_2022_battery}, fail when the differential index is higher than 1 because an increase in the differential index of DAEs introduces more difficulties in obtaining numerical solutions \citep{Petzold_1982_HighIndexDAE,Campbell_2008_DAE}. Many current DAE solvers are only capable of solving index-1 DAEs, so it is usually recommended to perform index reduction (e.g., \texttt{reduceDAEIndex} in MATLAB and \texttt{dae\_index\_lowering} in Julia) to transform a high-index system into an equivalent index-1 system, and then solve the resulting index-1 DAEs \citep{MATLAB_Index1DAE,Campbell_2008_DAE}. This technique is generally preferable because index-1 DAEs can be solved easily by most DAE solvers, eliminating the need for specialized high-index DAE solvers. Nevertheless, there are two major drawbacks associated with index reduction. First, the index reduction process is costly when the number of states and differential index are huge. Second, index reduction could introduce a large number of new variables, sometimes called dummy derivatives \citep{Mattsson_1993_dummy}, to replace high-order derivatives, which unnecessarily increases the complexity of the problem. These issues are not commonly discussed in the literature as they become significant when the differential index is sufficiently large, e.g., Problem 3 in our case studies. Thus, this approach does not match our objective of accelerating the optimal control algorithm.

An alternative to the index reduction approach is to fully discretize a system of high-index DAEs (e.g., using a collocation method) and solve the discretized equations \citep{Campbell_2008_DAE}. The specialized DAE solver in GEKKO employs the method of orthogonal collocation on finite elements to solve high-index DAEs \citep{GEKKO}. The main advantage of this implementation is that it does not entail any index reduction or dummy derivatives, making it more efficient when dealing with some high-index problems (e.g., see detailed discussions by \citet{APMonitor_2014} and \citet{GEKKO}). Hence, our DAE solutions for high-index systems rely on the DAE solver implemented in GEKKO.

\subsubsection{Proposed algorithm}
We denote this simulation-based approach as sim\_DAE, which consists of three main steps. First, replace the objective function with algebraic equations, reformulating the original optimal control problem as a system of DAEs 
\begin{equation} \label{eq:OptControlDAE}
    g(\Theta_1(\tau),\Theta_2(\tau),S(\tau),\Theta_b(\tau),\tau) = 0.
\end{equation}
The choice of algebraic equations is dependent on the objective function, which is demonstrated via the case studies presented in Section \ref{sec:CaseStudies}. 

Second, treat the control variable ($\Theta_b$ in this case) as a state instead of a decision variable. A consistent initial condition is required for this new state, and the initialization can be done in several ways, depending on the solver. For example, in GEKKO, the DAE solver does not strictly require a consistent initial condition, so it is possible to start with some reasonable guess and then run the DAE solver once to find a more accurate initial condition. Another possibility is to solve the optimization locally at the beginning of the process. Either approach requires minimal effort and computation, and therefore does not impact the overall computational performance. The initial condition for this new state is denoted by
\begin{equation} \label{eq:Initialization}
    \Theta_b(\tau_0) = \Theta_{b0}.
\end{equation}
After obtaining the DAEs and initial conditions, the final step is to solve the resulting DAEs \eqref{eq:OptControlDAE} and \eqref{eq:Initialization}. As the control is now treated as a state, the optimal control vector is obtained automatically from the DAE solver. 

This DAE reformulation technique inherently assumes that at least one of the inequality constraints, bounds, or algebraic equations resulting from reformulating the objective function is active. For problems in which this condition does not hold, the simulation-based approach is not guaranteed to produce an optimal solution, which is discussed in more detail in Section \ref{sec:problem3}. Changes in the active constraints or bounds can be incorporated in the simulation using a switching technique, with Section \ref{sec:problem4} providing a discussion that includes references to several examples.

\subsection{Implementation}
In this work, both optimization- and simulation-based techniques were implemented in MATLAB R2023a following the procedures introduced in Sections \ref{sec:Opt_based} and \ref{sec:Sim_based}, with GEKKO called and executed in Python 3.10. All simulations were performed on a computer equipped with an Intel\textsuperscript{\textregistered} Core\texttrademark\ 
 i9-13950HX processor (24 cores) and 128 GB RAM running 64-bit Windows 11. The choice and justification of important solver options are given in Appendix \ref{app:Solveroptions}.


\section{Case Studies} \label{sec:CaseStudies}
This section presents several optimal control problems for cell thawing to demonstrate the simulation-based technique and compare it with the optimization-based techniques. These examples are drawn from real problems and protocols associated with control and optimization of cell thawing and moving boundary problems. The case studies include both simple and complex problems to help illustrate the approach and assess the relationship between the comparative performance of the various algorithms and problem complexity.

\begin{figure*}
\centering
    \includegraphics[scale=.8]{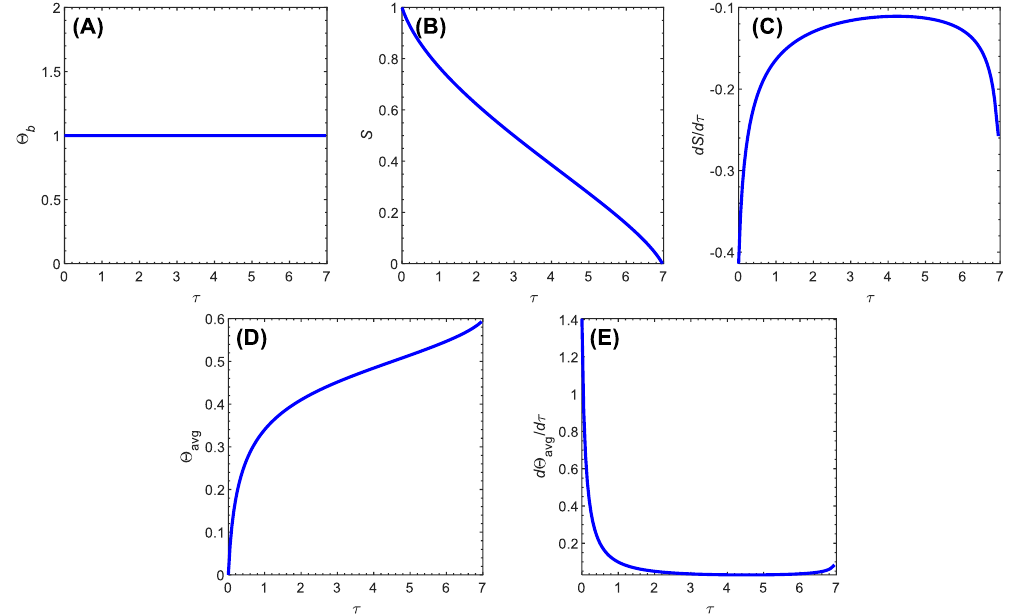}
   \vspace{-3pt}    
    \caption {Trajectories of the optimal (A) heater temperature $\Theta_b$, (B) interface position $S$, (C) interface velocity $dS/d\tau$, (D) average temperature $\Theta_\textrm{avg}$, and (E) rate of temperature change $d\Theta_\textrm{avg}/d\tau$ for Problem 1.} 
    \label{fig:Prob1_profile}      
\end{figure*}

\subsection{Problem 1: Minimization of the thawing time}
\subsubsection{Problem description and formulation}
This first problem is simple and intuitive, with the solution known {\it a priori}, to demonstrate the simulation-based technique and validate all algorithms.   

In general, it has been suggested that the heating process should be done rapidly to avoid potential damage to the cells and maintain high viability \citep{Terry_2010_RapidThawing,Baboo_2019_ThawingRates,Hunt_2019_TechnicalCondThawing,Uhrig_2022_ThawingOpt}, which is equivalent to the optimal control problem
\begin{equation} \label{eq:OptControl1}
\begin{aligned} 
& \min_{\Theta_b(\tau)} \quad  \tau_f \\
& \ \ \textrm{s.t.} 
    \ \textrm{Equations \eqref{eq:model1}--\eqref{eq:model8}, \eqref{eq:heatertemp_constraint}}. 
\end{aligned}
\end{equation}
The solution to this optimal control problem is straightforward. To minimize the thawing time, the heater temperature should be fixed at its upper bound to deliver the maximum heating power throughout the process. 

\subsubsection{DAE Reformulation}
To employ the simulation-based approach, \eqref{eq:OptControl1} needs to be reformulated as a DAE system. In this problem, the control vector $\Theta_b$ is treated as an algebraic state and explicitly specified an additional algebraic constraint $\Theta_b(\tau) = 1$. As a result, the equivalent system of DAEs for \eqref{eq:OptControl1} is
\begingroup
\allowdisplaybreaks
\begin{equation} \label{eq:OptControl1_DAE}
\begin{aligned}
    &\Theta_b(\tau) = 1, \\
    &\textrm{Equations \eqref{eq:model1}--\eqref{eq:model8}}. 
\end{aligned}
\end{equation}
\endgroup
The number of derivatives required to transform \eqref{eq:OptControl1_DAE} into an equivalent ODE system is 1, hence an index-1 DAE system. This index-1 DAE system can be solved easily by most DAE solvers.

\subsubsection{Solution comparison}
Every solution method gives the same optimal solution (Fig.~\ref{fig:Prob1_profile}); the heater temperature $\Theta_b$ is at its upper bound throughout the process, indicating that our algorithms and implementations are correct. The wall-clock time required for the simulation-based technique (sim\_DAE) is several orders of magnitude lower than those of the optimization-based approaches (Table \ref{Tab:ComputationalPerformance_Problem1}). Optimization with GEKKO is the slowest method, followed by IPOPT, CasADi and \texttt{fmincon}. GEKKO's ODE solvers cannot perform adaptive time-stepping, which could lead to significantly slower computation. Parallel computing fails to accelerate \texttt{fmincon} for this simple problem. Overall, the simulation-based approach is the most computationally efficient algorithm.

\begin{table}[ht]
\caption{Computational performance of each solution method for Problem 1.}
\renewcommand{\arraystretch}{1.2}
\label{Tab:ComputationalPerformance_Problem1}
\centering
\begin{tabular}{| l | c | c |}
\hline
\textbf{Method} & \textbf{Wall time (s)}\\ 
\hline
opt\_Ipopt &$10.74\pm0.22$\\
opt\_fmincon & $3.17 \pm 0.03$ \\
opt\_pfmincon & $3.23 \pm 0.06$ \\
opt\_sCasADi & $5.70 \pm 0.02$   \\
opt\_mCasADi & $5.81 \pm 0.06$  \\
opt\_Gekko & $34.08 \pm 0.31$ \\
sim\_DAE & $0.11 \pm 0.01$  \\
\hline
\end{tabular}
\renewcommand{\arraystretch}{1}
\end{table}

\subsection{Problem 2: Control of the average temperature}
\subsubsection{Problem description and formulation}
Instead of using the maximum temperature as in Problem 1, some past studies considered cases in which the freezing and thawing rates are controlled, i.e., the rate of temperature change is kept constant \citep{Shinsuke_2008_ControlledHeating,Jang_2017_CryopOverview,Baboo_2019_ThawingRates,Bojic_2021_Winter}. As such, this problem focuses on controlling the rate of temperature change during thawing by manipulating the heater temperature. The definition of the average temperature is as given in \eqref{eq:Tavg_numer}.

For a fixed heater temperature, the rate of temperature change is not constant (Figs.~\ref{fig:Prob1_profile}DE). Thus, the heater temperature needs to be manipulated. Consider the optimal control problem
\begingroup
\allowdisplaybreaks
\begin{equation} \label{eq:OptControl2}
\begin{aligned} 
& \min_{\Theta_b(\tau)} \ \  \bigintsss_{\tau_0}^{\tau_F} \!\!{\left( \dfrac{d\Theta_{\textrm{avg}}}{d\tau}-\left(\dfrac{d\Theta_{\textrm{avg}}}{d\tau}\right)_{\!\!\mathrm{sp}} \right)^{\!\!2} d\tau} \\
& \ \ \textrm{s.t.} \ \textrm{Equations \eqref{eq:model1}--\eqref{eq:model8}, \eqref{eq:heatertemp_constraint}},
\end{aligned}
\end{equation}
\endgroup
where $d\Theta_{\textrm{avg}}/d\tau$ is the rate of temperature change and $(d\Theta_{\textrm{avg}}/d\tau)_{\mathrm{sp}}$ is the target value (setpoint), set to 0.04 for demonstration in this problem. For the optimization-based approaches, the control vector is parameterized using a piecewise linear function, with the number of control intervals $n_c=16$ (Appendix \ref{app:ControlVector}).

\subsubsection{DAE Reformulation}
For the simulation-based technique, first analyze the objective function in \eqref{eq:OptControl2}. This objective function is minimized when the rate of temperature change is equal to the setpoint. Replacing the objective function in \eqref{eq:OptControl2} with the algebraic equation $d\Theta_{\textrm{avg}}/d\tau = (d\Theta_{\textrm{avg}}/d\tau)_{\textrm{sp}}$ results in the system of DAEs
\begingroup
\allowdisplaybreaks
\begin{equation} \label{eq:OptControl2_DAE}
\begin{aligned}
     &\dfrac{d}{d\tau} \!\!\left( \sum_{j=1}^{n} \dfrac{1}{n}{(\Theta_2)_j} \right) \!= \left(\dfrac{d\Theta_{\textrm{avg}}}{d\tau}\right)_{\!\!\mathrm{sp}}, \\
    &\textrm{Equations \eqref{eq:model1}--\eqref{eq:model8}, \eqref{eq:Initialization}}, 
\end{aligned}
\end{equation}
\endgroup
where $\Theta_b$ is now treated as a state. If the liquid domain used for calculating the average temperature includes the boundary point, the differential index of \eqref{eq:OptControl2_DAE} is 1. If the boundary point is excluded, the index is 2. In any cases, \eqref{eq:OptControl2_DAE} can be solved using GEKKO as explained in Section \ref{sec:Sim_based}.

\subsubsection{Solution comparison}
To evaluate the correctness of the solution, define the error measured by a modified 2-norm for this problem as 
\begin{equation} \label{eq:e2_Problem2}
    ||e||_2 = \sqrt{\dfrac{1}{n_k} \sum_{k=1}^{n_k}\left( \left(\dfrac{d\Theta_{\textrm{avg}}}{d\tau}\right)_{\!k}-\left(\dfrac{d\Theta_{\textrm{avg}}}{d\tau}\right)_{\!\!\mathrm{sp}} \right)^{\!\!2}},
\end{equation}
where $(d\Theta_{\textrm{avg}}/d\tau)_k$ is the rate of temperature change resulting from solving \eqref{eq:OptControl2} or \eqref{eq:OptControl2_DAE} evaluated at each time point $k$ in the time span [$\tau_0,\tau_f$] and $n_k$ is the number of sampling time points. A small value of $||e||_2$ corresponds to the rate of temperature change being close to the target value, indicating that the heater temperature is optimal; i.e., the solution method is accurate.

\begin{figure*}
    \centering
    \includegraphics[scale=.83]{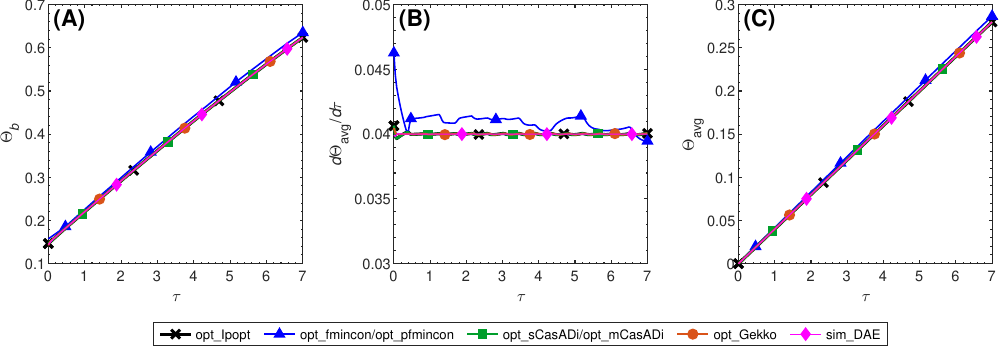}
       \vspace{-3pt}
    \caption{Trajectories of the optimal (A) heater temperature, (B) rate of temperature change, and (C) average temperature simulated via the optimization- and simulation-based approaches for Problem 2.} 
    \label{fig:Prob2_profile}
\end{figure*}

The optimal solutions to Problem 2 obtained from the optimization- and simulation-based approaches are illustrated in Fig.~\ref{fig:Prob2_profile}. Most solution techniques predict the same heater temperature profile except the methods with \texttt{fmincon}, which predict a slightly higher temperature (Fig.~\ref{fig:Prob2_profile}A). As time progresses, the heater temperature increases to compensate for a reduction in the temperature difference (driving force) between the heater and product. The rate of temperature change can be controlled nearly perfectly at 0.04 for IPOPT, CasADi, GEKKO, and the simulation-based method, while this value fluctuates up to about 0.046 for \texttt{fmincon}, implying that \texttt{fmincon} fails to converge to the optimal solution in this case (Fig.~\ref{fig:Prob2_profile}B). It is also evident that the average temperature increases linearly at the rate of 0.04 in all cases except for \texttt{fmincon} (Fig.~\ref{fig:Prob2_profile}C).

\begin{table}[ht]
\caption{Computational performance and accuracy of each solution method for Problem 2.}
\renewcommand{\arraystretch}{1.2}
\label{Tab:ComputationalPerformance_Problem2}
\centering
\begin{tabular}{| l | c | c |}
\hline
\textbf{Method} & \textbf{Wall time (s)} & \boldmath$||e||_2$\\ 
\hline
opt\_Ipopt & $162.67\pm3.72$  & 8.03$\times10^{-5}$  \\
opt\_fmincon & $349.07\pm11.67$  & 1.21$\times10^{-3}$  \\
opt\_pfmincon & $79.40\pm1.57$  & 1.21$\times10^{-3}$  \\
opt\_sCasADi & $363.36\pm12.52$  & 8.04$\times10^{-5}$  \\
opt\_mCasADi & $5189.15\pm71.42$  & 8.04$\times10^{-5}$  \\
opt\_Gekko & $191.89\pm4.86$  & 4.91$\times10^{-5}$  \\
sim\_DAE & $3.71\pm0.15$  & 2.25$\times10^{-5}$  \\
\hline
\end{tabular}
\renewcommand{\arraystretch}{1}
\end{table}

The simulation-based approach is by far the fastest solution method, accelerating the computation by 40$\times$ to 90$\times$ compared to the typical solvers IPOPT and \texttt{fmincon} (Table \ref{Tab:ComputationalPerformance_Problem2}). Without parallel computing, optimization with IPOPT is fastest, followed by GEKKO, \texttt{fmincon}, and CasADi. Parallel computing can significantly reduce the computation time for \texttt{fmincon}. With the direct multiple shooting method (opt\_mCasADi), the number of states resulting from spatial discretization is large. For example, with $n=20$, there are 41 states for each control interval and 656 states for $n_c = 16$. This large nonlinear optimization is computationally expensive to solve. In terms of accuracy, \texttt{fmincon} produces the largest error, which is about 30--40 fold higher than for the other techniques. This indicates that the solution obtained from \texttt{fmincon} is not optimal, agreeing with the plots shown in Fig.~\ref{fig:Prob2_profile}. Although 
 both \texttt{fmincon} and IPOPT employ the interior point methods, the detailed implementations (e.g., gradient/hessian calculation, scaling) are not identical, which could result in different accuracy and performance (see detailed discussion and comparison of these solvers in
\citet{Rojas-Labanda_2015_opt}). The simulation-based and other optimization-based methods except \texttt{fmincon} have similar accuracy. In this problem, the solution is now known {\it a priori} as in Problem 1, but the simulation-based approach is still much faster than the optimization-based techniques while giving the same level of accuracy. 

\subsection{Problem 3: Control of the interface velocity}
\label{sec:problem3}

\subsubsection{Problem description and formulation}
One of the primary interests in a Stefan problem is the moving solid-liquid interface. The literature has studied and shown the importance of interface tracking and optimization in various applications \citep[for example]{Hill_1994_CastingInterface,Brezina_2018_PCM,Srisuma_2023_1DCellThawingEstimation}. This problem focuses on controlling the interface velocity, i.e., the evolution of a melting/freezing process.

Consider the optimal control problem
\begin{equation} \label{eq:OptControl3}
\begin{aligned} 
& \min_{\Theta_b(\tau)} \ \   \bigintsss_{\tau_0}^{\tau_F} \!{\left( \dfrac{dS}{d\tau}-\left(\dfrac{dS}{d\tau}\right)_{\!\!\mathrm{sp}} \right)^{\!\!2} d\tau} \\
& \ \ \textrm{s.t.} \ \textrm{Equations \eqref{eq:model1}--\eqref{eq:model8}, \eqref{eq:heatertemp_constraint}},\end{aligned}
\end{equation}
where $dS/d\tau$ is the interface velocity and $\left(dS/d\tau\right)_{\mathrm{sp}}$ is the target velocity, set to $-0.1$ for demonstration in this problem. In Problem 1 (Figs.~\ref{fig:Prob1_profile}BC), the interface velocity is not constant when the heater temperature is fixed. In this problem, the aim is to control the interface velocity to be constant at $(dS/d\tau)_\mathrm{sp}$ by manipulating the heater temperature throughout the process as formulated in \eqref{eq:OptControl3}. For the optimization-based approaches, the control vector is parameterized using a piecewise linear function, with the number of control intervals $n_c = 12$ (see Appendix \ref{app:ControlVector}).

\subsubsection{DAE Reformulation}
To apply the simulation-based technique, a similar approach used in Problem 2 can be used, but here we instead enforce the interface velocity by modifying the interface equation \eqref{eq:model5} to be equal to $(dS/d\tau)_\textrm{sp}$. Consequently, \eqref{eq:OptControl3} is reformulated as the DAE system 
\begingroup
\allowdisplaybreaks
\begin{equation}  \label{eq:OptControl3_DAE}
\begin{aligned}
    & \dfrac{dS}{d\tau} = f_3(\Theta_1, \Theta_2, S) = \left(\dfrac{dS}{d\tau}\right)_{\!\!\mathrm{sp}}, \\
    &\textrm{Equations \eqref{eq:model1}, \eqref{eq:model2}, \eqref{eq:model6}--\eqref{eq:model8}, \eqref{eq:Initialization}}. 
\end{aligned}
\end{equation}
\endgroup
The differential index of \eqref{eq:OptControl3_DAE} is $n$, which is the number of grid points resulting from spatial discretization. Therefore, if the spatial discretization of the domain is made finer to increase numerical accuracy, the differential index of \eqref{eq:OptControl3_DAE} also increases. In this case, $n=20$, and hence \eqref{eq:OptControl3_DAE} becomes an index-20 DAE system, making the problem even more complicated and difficult to solve than Problems 1 and 2. 

To explore the possibility of using the index reduction technique, MATLAB's \texttt{reduceDAEIndex} was used to reduce the differential index of \eqref{eq:OptControl3_DAE}. However, the index reduction process was not complete even after 20 hours of simulation. Obviously, this technique is not feasible for our problem, as explained before in Section \ref{sec:Sim_based}. Therefore, GEKKO's DAE solver is needed.

\subsubsection{Solution comparison}
Similar to Problem 2, we define the error measured by a modified 2-norm for this problem as 
    \begin{equation} \label{eq:e2_Problem3}
        ||e||_2 = \sqrt{\dfrac{1}{n_k} \sum_{k=1}^{n_k}\left( \left(\dfrac{dS}{d\tau}\right)_{\!\!k}-\left(\dfrac{dS}{d\tau}\right)_{\!\!\mathrm{sp}} \right)^{\!\!2}},
    \end{equation}
where $(dS/d\tau)_k$ is the interface velocity resulting from solving \eqref{eq:OptControl3} or \eqref{eq:OptControl3_DAE} evaluated at each time point $k$ in the time span [$\tau_0,\tau_f$].

\begin{figure*}[ht]
    \centering
    \includegraphics[scale=.83]{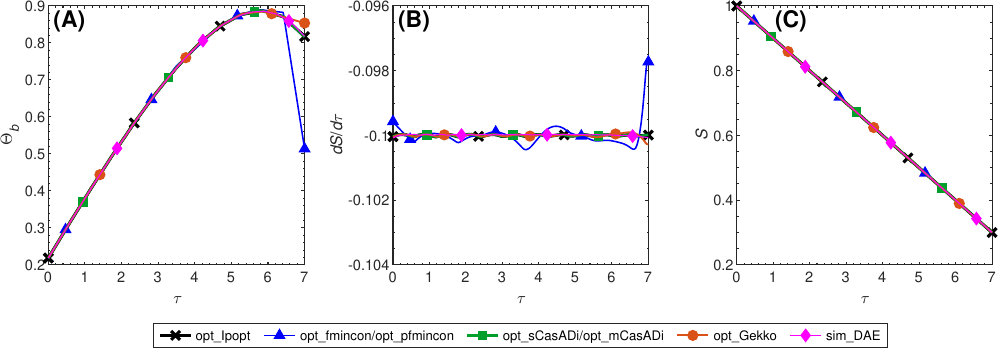} 
       \vspace{-3pt}
    \caption{Trajectories of the optimal (A) heater temperature, (B) interface velocity, and (C) interface position via the optimization- and simulation-based approaches for Problem 3.} 
    \label{fig:Prob3_profile}
\end{figure*}

The optimal solutions to Problem 3 are shown in Fig.~\ref{fig:Prob3_profile}. The optimal heater temperature predicted by each technique is nearly identical except that there is a large drop near the end of the process for \texttt{fmincon} (Fig.~\ref{fig:Prob3_profile}A). The interface velocity can be controlled at about $-0.1$ throughout the process for IPOPT, CasADi, GEKKO, and the simulation-based technique, while some fluctuation is observed for \texttt{fmincon} (Fig.~\ref{fig:Prob3_profile}B). This indicates that \texttt{fmincon} is less accurate than the others approaches, which is a similar trend observed before in Problem 2. With the interface velocity controlled, the interface position recedes linearly at the rate of 0.1 (Fig.~\ref{fig:Prob3_profile}C).

For the computational performance, the simulation-based technique is about 60$\times$ to 900$\times$ faster than all of the optimization-based approaches (Table \ref{Tab:ComputationalPerformance_Problem3}), despite an increase in the differential index. IPOPT, GEKKO, and CasADi are comparable for the direct single shooting method, while the multiple shooting method is much slower. Parallel computing can significantly accelerate \texttt{fmincon} but is still much slower than the simulation-based approach. In terms of accuracy, the error of the simulation-based approach is about $2$$\times$$10^{-5}$, which is similar to those of the optimization-based approaches except for \texttt{fmincon}. With a relatively large error, \texttt{fmincon} is the least accurate method in this problem. 

\begin{table}[ht]
\caption{Computational performance and accuracy of each solution method for Problem 3.}
\renewcommand{\arraystretch}{1.2}
\label{Tab:ComputationalPerformance_Problem3}
\centering
\begin{tabular}{| l | c | c |}
\hline
\textbf{Method} & \textbf{Wall time (s)} & \boldmath$||e||_2$\\ 
\hline
opt\_Ipopt & $214.01\pm5.75$ & 1.96$\times10^{-5}$  \\
opt\_fmincon & $317.89\pm5.46$ & 3.05$\times10^{-4}$  \\
opt\_pfmincon & $92.30\pm1.52$ & 3.05$\times10^{-4}$  \\
opt\_sCasADi & $225.42\pm5.92 $ & 1.96$\times10^{-5}$  \\
opt\_mCasADi & $2618.13\pm83.29$ & 1.97$\times10^{-5}$  \\
opt\_Gekko & $179.77\pm2.45$ & 4.37$\times10^{-5}$  \\
sim\_DAE & $2.99\pm0.07$ & 2.41$\times10^{-5}$  \\
\hline
\end{tabular}
\renewcommand{\arraystretch}{1}
\end{table}

From the three case studies, the simulation-based approach reliably solves the optimal control problems irrespective of the differential index, ranging from a simple index-1 problem to an extremely high-index DAE system. The approach is much faster than any optimization-based technique while giving the same level of accuracy. 

The simulation-based approach, however, has one limitation associated with the DAE reformulation. As described in Section \ref{sec:Sim_based}, the DAE reformulation is guaranteed to be optimal when at least one of the inequality constraints, bounds, or algebraic equations resulting from reformulating the objective function is active. This is typically the case when the objective function of the original optimal control problem can be minimized by solving each subproblem resulting from control vector parameterization individually -- in other words, local minimization at every control/time interval (subproblem) is equivalent to global minimization of the whole process (original problem). For example, in Problem 1, if the thawing time of every control interval is minimized, the total thawing time is automatically minimized. In Problems 2 and 3, if the rate of temperature change and interface velocity are controlled locally for each control interval, the whole process is also controlled. These objective functions are common in many applications, for which the simulation-based technique can be applied, including lithium-ion batteries of various chemistries \citep{Berliner_2022_battery,Galuppini2023Battery,Matschek2023Battery} and microwave lyophilization \citep{Srisuma2023Analytical}. An example of a system in which the method is not optimal is a reactor with multiple chemical reactions, with the goal of maximizing the amount of desired product at the end of the process. Maximizing the amount of product during the first control interval could concurrently increase the amount of some byproducts that can degrade the desired product later. In this situation, local optimization is not equivalent to global optimization. As such, an algebraic equation/constraint may not be active, so an optimization-based technique is needed. Although the simulation-based approach does not give an optimal solution to such problems, it can be used to provide an initialization to the optimization solver.

\subsection{Problem 4: Sensitivity analysis}
\label{sec:problem4}

In Problems 2 and 3, the simulation-based approach is tested with one set of values of $(d\Theta_{\textrm{avg}}/d\tau)_{\mathrm{sp}}$ and $(dS/d\tau)_\textrm{sp}$, which are 0.04 and $-0.1$, respectively. To demonstrate the robustness of our framework, this section conducts a sensitivity analysis by varying $(d\Theta_{\textrm{avg}}/d\tau)_{\mathrm{sp}}$ and $(dS/d\tau)_\textrm{sp}$ and employs the simulation-based approach to solve the problems.

\begin{figure*}
    \centering
    \includegraphics[scale=.83]{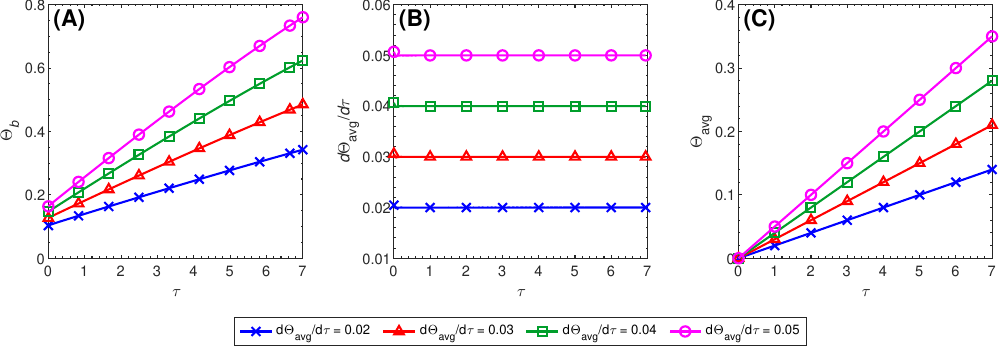}
    
   \vspace{-3pt}
    
    \caption{Trajectories of the optimal (A) heater temperature, (B) rate of temperature change, and (C) average temperature simulated via the simulation-based approach at four different values of $(d\Theta_{\textrm{avg}}/d\tau)_{\mathrm{sp}}$ for Problem 4.} 
    \label{fig:Prob4_temp}
\end{figure*}

\begin{figure*}
    \centering
    \includegraphics[scale=.83]{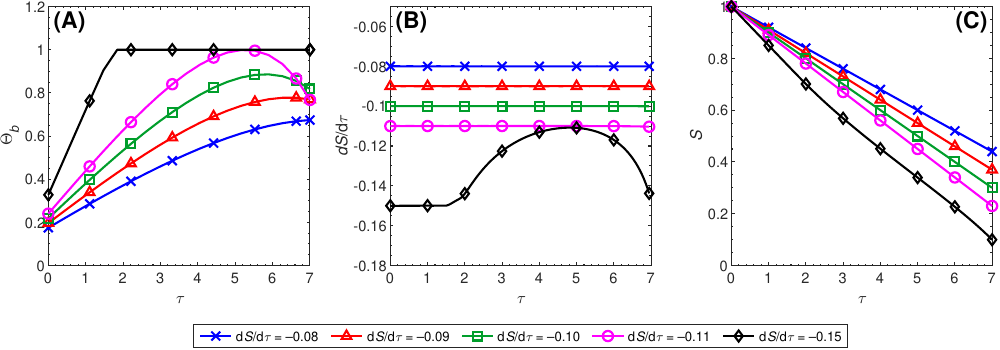}
    \vspace{-3pt}
    \caption{Trajectories of the optimal (A) heater temperature, (B) interface velocity, and (C) interface position simulated via the simulation-based approach at four different values of $(dS/d\tau)_\textrm{sp}$ for Problem 4.} 
    \label{fig:Prob4_Itf}
\end{figure*}

The simulation-based approach provides accurate solutions irrespective of the values of $(d\Theta_{\textrm{avg}}/d\tau)_{\mathrm{sp}}$ and $(dS/d\tau)_\textrm{sp}$ (Figs.~\ref{fig:Prob4_temp} and \ref{fig:Prob4_Itf}). The rate of temperature change and interface velocity are at the setpoints except for $(dS/d\tau)_\textrm{sp}=-0.15$. An interface velocity of $-0.15$ is too fast for the given heater temperature, so the upper bound is active, and that the heater temperature cannot be increased further to achieve the target velocity. A target velocity that is larger (smaller in magnitude) than the peak of Fig.~\ref{fig:Prob1_profile}C, which about $-0.12$, will never violate the bound (see Appendix \ref{app:bound} for a formal proof).

For cases where the bound is active, a simulation-switching technique can be used to transition between active constraints, resulting in a hybrid discrete/continuous dynamic simulation \citep{Feehery1998Switching,Barton2000Hybrid,Berliner_2022_battery}. First, initialize a DAE simulation as usual. Next, terminate the current simulation when the bound on any control vector is active and then initialize a new DAE simulation with that active bound/constraint. Perform this switching whenever there is a change in the active constraints, which results in a mixed continuous-discrete DAE system instead of a pure continuous system. The complexity of this procedure depends on the choice of a DAE solver. For example, MATLAB's and Julia's DAE solvers have a built-in function to track all variables and terminate a simulation when the specified condition (aka event) is met, which facilitates the implementation of such switching. The switching technique can also be used to handle cases where the control trajectory is discontinuous (see examples in \citet{Berliner_2022_battery,Srisuma2023Analytical}). GEKKO's DAE solvers, on the other hand, do not have a built-in option for handling switches, and so this process needs to be executed manually. This switching technique can also be extended to handle path constraints.

With this capability, the benefit of the simulation-based approach is even more substantial in applications where a large number of simulation runs are required, e.g., parametric studies and design optimization.

\section{Conclusion} \label{sec:Conclusion}
This article describes a new approach for determining control and state trajectories in optimal control problems by reformulation as a system of DAEs. The optimal control and state vectors are obtained via simulation of the resulting DAE system with the selected DAE solver, eliminating the need for an optimization solver and thus greatly accelerating the computation. 

Our proposed framework, the simulation-based approach, is demonstrated and benchmarked against a variety of optimization-based approaches -- IPOPT, \texttt{fmincon}, CasADi, and GEKKO -- for case studies involving the optimal control of a Stefan problem for cell thawing. The case studies include the minimization of the thawing time, the control of the average temperature, and the control of the interface velocity, covering both index-1 and high-index DAE systems. The simulation-based approach is faster than every optimization-based method by more than an order of magnitude while giving similar/better accuracy in all cases. The current simulation-based approach, however, is limited to problems where at least one of the inequality constraints, bounds, or algebraic equations resulting from reformulating the objective function is active. 

Some future directions of interest are to (1) generalize the approach for more complicated problems, objective functions, and constraints that arise in multiple fields and (2) improve the computational performance with more efficient solvers and implementation.

\section*{Acknowledgments} 
The authors would like to thank Dr. John D. Hedengren for advice on the software GEKKO.

\section*{Appendices}
\appendix

\section{Solver Options} \label{app:Solveroptions}
There are a large number of options for optimization, ODE, and DAE solvers, which could lead to different accuracy and computational performance. This section describes and justifies the choice of important solver options used in our simulations. The default values are used for options that are not listed here.

For all case studies, the selected DAE solvers are MATLAB's \texttt{ode15s} for index-1 DAEs and GEKKO's DAE solver (Python) for high-index systems. The differential index of DAEs can be checked using MATLAB's \texttt{reduceDAEIndex}. The wall-clock times (aka wall times, clock times) can be measured using the \texttt{tic} and \texttt{toc} functions, with each simulation repeated at least 10 times until the standard deviation of the measured wall times is smaller than 5\% for consistent results. The initial guess of $\Theta_b$ is set to 0.5, the average dimensionless temperature, so that every solution method is initialized and compared on the same basis. 

For the optimization-based approaches, the optimality tolerance is set to $10^{-7}$ for all optimization solvers, including IPOPT, fmincon, CasADi, and GEKKO. The integration tolerance for \texttt{ode15s} is $10^{-10}$. Using higher values for the optimality and integration tolerances could lead to inaccurate solutions, while using tighter values does not improve the accuracy significantly. For the objective functions \eqref{eq:OptControl2} and \eqref{eq:OptControl3} in Problems 2 and 3, the derivatives are approximated using a finite difference scheme, whereas the integrals can be approximated using a Riemann sum. The time step $\Delta\tau$ is set to 0.05. 

While \texttt{ode15s} uses adaptive time-stepping, this option is not available in GEKKO; i.e., users manually specify a vector of time points for ODE integration. Hence, there is no ideal comparison between GEKKO's ODE solvers and MATLAB's \texttt{ode15s}. The fixed time step $\Delta\tau=0.05$ is specified for ODE integration in GEKKO, with five collocation points (INODES = 5) for each time interval. This time step is chosen to be consistent with the value used for a finite difference and Riemann sum approximation mentioned in the previous paragraph, while the number of collocation points is selected such that that the temperature and interface position simulated by GEKKO's ODE solver have the same level of accuracy as those simulated by \texttt{ode15s}. 

For the simulation-based approach, \texttt{ode15s} is used for an index-1 system (Problem 1), with the exact same solver options as mentioned above. GEKKO's DAE solver is used for high-index systems (Problems 2 and 3). As a reformulated DAE system is different from the original ODE system, the time step needs to be adjusted accordingly. The time step is selected such that the accuracy of the simulation-based solution measured by $||e||_2$ is on the same order of magnitude as that of the optimization-based solution, giving $\Delta\tau=0.12$ and $\Delta\tau=0.37$ for Problems 2 and 3, respectively.   

The accuracy of the optimization- and simulation-based approaches is measured using $||e||_2$ defined by \eqref{eq:e2_Problem2} and \eqref{eq:e2_Problem3}. To calculate $||e||_2$, the derivatives are approximated using a finite difference scheme with $\Delta\tau=0.05$.

\section{Control Vector Parameterization} \label{app:ControlVector}
This appendix describes the choice of control vector parameterization used for the optimization-based approaches.

\subsection{Piecewise constant and linear controls}
Piecewise constant and linear controls are most common in optimal control \citep{Nolasco_2020_OptimalControl}. This section justifies the use of piecewise linear controls in Problems 2 and 3 for the optimization-based approaches, in comparison to piecewise constant controls. This comparison considers the most complex problem, Problem 3, and uses opt\_Ipopt as the solver. 

\begin{table}[ht]
\caption{Comparison between the errors $||e||_2$ of piecewise constant and linear control vector parameterization for different values of control intervals $n_c$.}
\renewcommand{\arraystretch}{1.2}
\label{Tab:PW_Con-Lin}
\centering
\begin{tabular}{| c | c | c |}
\hline
\multirow{2}{*}{\textbf{$n_c$}} & \multicolumn{2}{c|}{Error measured by $||e||_2$} \\ \cline{2-3}
&  Piecewise constant & Piecewise linear \\
\hline
4 & 0.012 & 3.01$\times$$10^{-4}$ \\
8 & 0.010 & 5.68$\times$$10^{-5}$ \\
12 & 0.012 & 1.96$\times$$10^{-5}$ \\
\hline
\end{tabular}
\renewcommand{\arraystretch}{1}
\end{table}

From Table \ref{Tab:PW_Con-Lin}, the piecewise linear control is more accurate than the piecewise constant control by many orders of magnitude. Since the dynamics of the interface position and average temperature are highly nonlinear, a large value of $n_c$ is required for a piecewise constant control to accurately manipulate the average temperature and moving interface, leading to a much larger nonlinear program that is computationally expensive and difficult to converge. As a result, a piecewise linear control is selected in this work.

\subsection{Number of control intervals}
The accuracy of the optimization-based solution is influenced by the number of control intervals $n_c$. This section justifies those numbers by investigating the effect of $n_c$ on the accuracy of the solution.

\begin{figure}[ht]
    \centering
    \includegraphics[scale=.95]{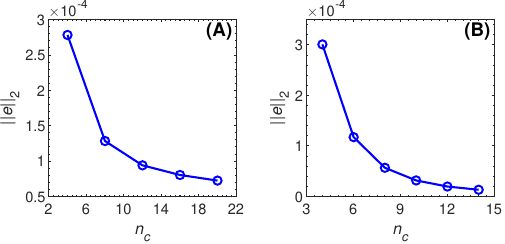}
    \vspace{-10 pt}
    \caption{Errors measured by $||e||_2$ for different values of control intervals $n_c$ for (A) Problem 2 and (B) Problem 3.} 
    \label{fig:Control_Interval}
\end{figure}

The optimal control problems defined by \eqref{eq:OptControl2} and \eqref{eq:OptControl3} are solved via opt\_Ipopt and $||e||_2$ is calculated for different values of $n_c$. Logically, the accuracy of the solution should improve (i.e., $||e||_2$ decreases) when $n_c$ increases, due to finer discretization. From Fig.~\ref{fig:Control_Interval}, the accuracy of the solution does not change significantly after $n_c$ reaches some certain value, and so this threshold should be chosen for control vector parameterization. As a result, $n_c$ is set to 16 for Problem 2 (Fig.~ \ref{fig:Control_Interval}A) and 12 for Problem 3 (Fig.~\ref{fig:Control_Interval}B).

\section{Bounds on the Interface Velocity and Rate of Temperature Change} \label{app:bound}
We show in Fig.~\ref{fig:Prob1_profile}, Problems 2 and 3, and Section 4.4 that, if the target velocity (and also the rate of temperature change) is chosen properly, the bound on a control vector will never be violated. A formal proof is given in this section.

To obtain a closed-form solution for the proof, heat conduction in the liquid domain is assumed to be quasi-steady. This approximation is accurate for problems associated with phase change in general as the sensible heat is much smaller than the latent heat. In this case, the closed-form solution for the interface velocity is
\begin{equation}\label{eq:dSdt}
    \frac{dS}{d\tau} = -\frac{k_2U\Theta_b(T_0-T_m)}{\rho L\alpha_1S\left(U\ln(1/S) + k/b\right)},
\end{equation}
where the parameter values and description can be found in \citet{Srisuma_2023_1DCellThawingEstimation}. From \eqref{eq:dSdt}, the interface velocity $dS/d\tau$ is a monotonic function of the heater temperature $\Theta_b$. Substituting the upper bound $\Theta_b = 1$ into \eqref{eq:dSdt} yields
\begin{equation}\label{eq:bound_dSdt}
    \left(\dfrac{dS}{d\tau}\right)_{\!\!\mathrm{sp}} > -\frac{k_2U(T_0-T_m)}{\rho\alpha_1 \Delta H_{\!f}S\left(U\ln(1/S) + k/b\right)},
\end{equation}
which gives the limit on the target velocity that can be specified without violating the upper bound on the heater temperature.
With the parameter values in \citet{Srisuma_2023_1DCellThawingEstimation}, the right-hand side of \eqref{eq:bound_dSdt} is about $-0.13$. This conclusion agrees with the results presented in Problems 3 and 4, in which the bounds are not violated when $(dS/d\tau)_{\mathrm{sp}}$ is higher than $-0.13$, whereas the upper bound is active when $(dS/d\tau)_{\mathrm{sp}}$ is $-0.15$. A similar strategy can be used to analyze the temperature change, e.g., using a thermal lumped capacity approximation.

Note that our simulation-based approach does not require to know {\it a priori} if the bounds will be active or not, as the upper and lower bounds can be handled as shown in Problem 4.

\section*{Data Availability}  \label{sec:Code}
Software and data used in this work are available at \url{https://github.com/PrakitrSrisuma/simDAE-optimalcontrol}. 

\bibliographystyle{elsarticle-harv}
\bibliography{reference}  

\pagebreak
\begin{wrapfigure}{l}{25mm} 
    \includegraphics[width=1in,height=1.25in,clip,keepaspectratio]{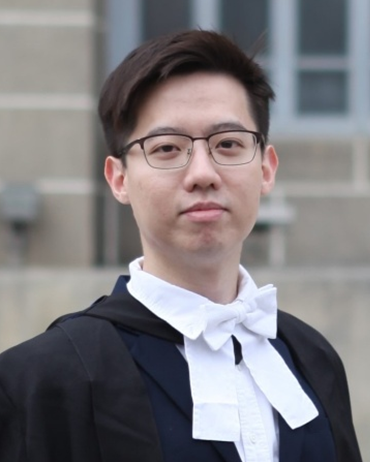}
\end{wrapfigure}\par
\textbf{Prakitr Srisuma} is a PhD candidate in the group of Prof. Richard D. Braatz at the Massachusetts Institute of Technology (MIT). He received an MPhil in Advanced Chemical Engineering from the University of Cambridge and a BEng in Chemical Engineering from Chulalongkorn University. Before moving to MIT, he was a process engineer at PTT Exploration and Production in Thailand. His current research is in process systems engineering for biopharmaceutical manufacturing, focusing on mathematical modeling and model-based control. Recognitions include the AIChE PD2M Student Award and Thai Graduate Fulbright Scholarship.\par

\begin{wrapfigure}{l}{25mm} 
    \includegraphics[width=1in,height=1.25in,clip,keepaspectratio]{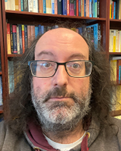}
\end{wrapfigure}\par
\textbf{George Barbastathis} is a Professor of Mechanical Engineering at the Massachusetts Institute of Technology (MIT), where he does research in machine learning and optimization for computational imaging and inverse problems; and in optical physics, including statistical optics, scattering theory, and artificial optical materials and interfaces. He received the Diploma in Electrical and Computer Engineering in 1993 from the National Technical University of Athens and an MSc and PhD in Electrical Engineering in 1994 and 1997, respectively, from the California Institute of Technology. He is a member of the Institute of Electrical and Electronics Engineering (IEEE) and the American Mathematical Society (AMS), a Fellow of the Optical Society of America (OSA), which was recently rebranded as Optica, and a Fellow of the Society for Photo Instrumentation Engineering (SPIE). \par

\begin{wrapfigure}{l}{25mm} 
    \includegraphics[width=1in,height=1.25in,clip,keepaspectratio]{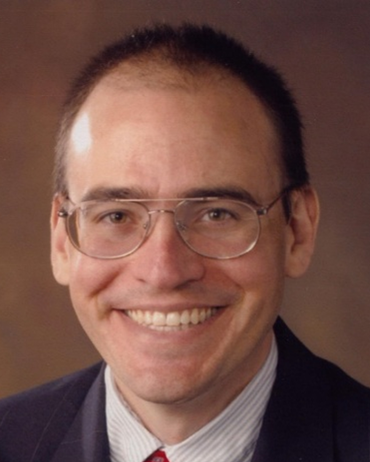}
\end{wrapfigure}\par
\textbf{Richard D. Braatz} is the Edwin R. Gilliland Professor at the Massachusetts Institute of Technology (MIT), where he does research in control theory and its applications to advanced manufacturing. He received an MS and PhD from the California Institute of Technology and was a Professor at the University of Illinois at Urbana-Champaign and a Visiting Scholar at Harvard University before moving to MIT. Recognitions include the AACC Donald P. Eckman Award and the IEEE CSS Antonio Ruberti Young Researcher Prize. He is a Fellow of IEEE and IFAC and a member of the U.S. National Academy of Engineering. \par

\end{document}